\documentclass[12pt,authoryear,review]{elsarticle}

\makeatletter
\def\ps@pprintTitle{%
 \let\@oddhead\@empty
 \let\@evenhead\@empty
 \def\@oddfoot{}%
 \let\@evenfoot\@oddfoot}
\makeatother

\usepackage{dsfont}
\usepackage{epstopdf}
\usepackage{mathrsfs}
\usepackage{amsfonts}
\usepackage{amssymb}
\usepackage{amsmath}

\usepackage{graphicx}
\usepackage{color}
\usepackage{extarrows}
\usepackage{float}
\usepackage{bigstrut,multirow,rotating}

\newcommand{\EE}{{\mathbb E}}
\newcommand{\VV}{{\mathbb V}}

\newcommand{\PP}{{\mathbb P}}

\newcommand{\Ft}{{\mathcal F}}

\newcommand{\blambda}{{\boldsymbol{\lambda}}}
\newcommand{\bpi}{{\boldsymbol{\pi}}}
\newcommand{\bP}{{\boldsymbol{P}}}

\newcommand{\bA}{{\boldsymbol{A}}}

\newcommand{\bG}{{\boldsymbol{G}}}

\usepackage{color}

\newcommand{\hV}{{\vartheta}}

\newproof{pf}{Proof}

\begin{document}
	\begin{frontmatter}
		
		\title {\textbf{Technical Uncertainty in Real Options with Learning}\tnoteref{t1}\\[0.1em]
		{\normalsize \textbf{The Journal of Energy Markets, Forthcoming}}}		
		\tnotetext[t1]{We are grateful to two anonymous referees for valuable comments.}				
		
		\author[author1] {Ali Al-Aradi}
		\ead{ali.al.aradi@utoronto.ca}
		\author[author2]{\'Alvaro Cartea}
		\ead{Alvaro.Cartea@maths.ox.ac.uk}
		\author[author3]{Sebastian Jaimungal}
		\ead{sebastian.jaimungal@utoronto.ca}
				\address[author1] {Department of Statistical Sciences, University of Toronto, Toronto, Canada}
		\address[author2] {Mathematical Institute, University of Oxford, Oxford, UK\\  Oxford-Man Institute of Quantitative Finance, Oxford, UK}
		\address[author3] {Department of Statistical Sciences, University of Toronto, Toronto, Canada}

		\begin{abstract}
	 We introduce a new approach to incorporate uncertainty into the decision to invest in a commodity reserve.  The investment is an irreversible one-off capital expenditure,  after which the investor receives a stream of cashflow from extracting the commodity and selling it on the spot market. The investor is  exposed to  price uncertainty and uncertainty in the amount of available resources in the reserves (i.e. technical uncertainty). She does, however, learn about the reserve levels through time, which is a key determinant in the decision to invest. To model  the reserve level uncertainty and how she learns about the  estimates of the commodity in the reserve, we adopt a continuous-time Markov chain model to value the option to invest in the reserve and  investigate the value that learning has prior to investment.
	\end{abstract}
		
		
	\end{frontmatter}
\section{Introduction}

\vspace{-0.5cm}

What are the optimal market conditions to make an investment decision is  an extensively studied  problem in the academic literature and a key question at the heart of the valuation and execution of projects under uncertainty. Some investment projects are endowed with the option to delay decisions until market conditions are optimal. This option is valuable because decisions are made when the potential gains stemming from the decision are maximized.  The classical work of \cite{McDonaldSiegel86} is the first to formalize the investment problem as a real option to invest in a project. In their work, the value $O_t$ of the option  is calculated by comparing the difference in the value of investing now and the value of making the investment at a future time. Specifically, the value of the real option is
\begin{equation}
O_t = \sup_{\tau \in \mathcal{T}} \EE\left[e^{-\rho(\tau-t)}(V_{\tau} - I_{\tau})_+\right]\,,
\end{equation}
where $\mathcal{T}$ is the set of admissible $\Ft$-stopping (exercise) times, $\rho$ is the risk-adjusted discount rate, and $V_t$ and $I_t$ are the project value and (sunk) cost respectively. The project value and cost are traditionally modeled with a geometric Brownian motion (GBM). The solution to this problem shows that the optimal investment strategy is to invest when the ratio ${V_t}/{I_t}$ reaches a critical boundary $B$ (the problem of optimal scrapping/divesting is similar, with the roles of $V_t$ and $I_t$ reversed). 
More recently, several authors have studied this problem with mean-reverting project value and costs (see, e.g., \cite{MetcalfHasset95}, \cite{Sarkar03}, \cite{JaiSouzaZubelli09}).

In another classical paper, \cite{BrennanSchwartz1985} focus on the management of a mine (controlling output rate, opening/closing of mine, abandonment, and so on) rather than the optimal timing problem (see also \cite{Dixit89}). Management decisions are modulated by the output prices which are modeled as a GBM, while costs are known.


These classical works do not take into account the uncertainty associated with reserve levels. To account for such ``technical uncertainty'', \cite{Pindyck80} develops a model where the demand and reserve levels fluctuate continuously with increasing variance. Furthermore, the optimal strategy is influenced by exploration and is introduced as a policy (i.e. control) variable in two distinct ways. The first allows exploratory effort to affect the level of ``knowledge'', which reduces the variance of reserve fluctuations.  The second assumes that reserves are discovered at a rate that depends on: how much has already been discovered in the past, the amount of current effort, and exogenous noise.

More recent approaches to the investment timing problem with technical uncertainty include those using Bayesian updating as in \cite{ArmstrongGalliBaileyCouet}, modeling project costs via Markov chains as in \cite{ElliotMiaoYu2009}, and using proportionality to model learning as in \cite{Sadowsky2005}. Also, \cite{CortozarSchwartzCasassus2000} describes a comprehensive approach to valuing several-stage exploration, solving the timing problem, and provides investment management (closure, opening, etc.) decision rules -- see also \cite{BrennanSchwartz1985}. Other works that incorporate real option techniques in the valuation of flexibility and investment decisions in commodities and energy include that of \cite{himpler2014optimal} who look at the optimal timing of wind farm repowering;  \cite{taschini2010real} who look at the option to switch fuels under different scenarios and fuel incentives; the work of \cite{fleten2011transmission} that looks at the option to choose the capacity of an electricity interconnector between two locations, and \cite{cartea2012much} who value an electricity interconnector as a stream of real options of the difference of prices in two locations. Meanwhile, \cite{cartea2017irreversible} study the effect that model uncertainty has on irreversible investments.

This work adds to the literature by incorporating both market and reserve uncertainty, while allowing the agent to learn about the status of reserves. Reserve uncertainty is represented by a Markov chain model with transition rates that decay as time flows forward to mimic the notion of learning. The model setup is developed in the context of oil exploration, however, it may be applied to other investment problems in commodities, such as mining for precious or base metals,  and natural gas fields.  We value the irreversible option to invest in the exploration by developing a version of Fourier space-time stepping, as in \cite{Jai08}, \cite{Surkov11}, and \cite{jaimungal2013valuing} for equity, commodity and interest-rate derivatives, respectively. For other work on ambiguity aversion and model uncertainty in commodities and algorithmic trading see \cite{CJZ}, \cite{CDJ},  \cite{Dempster},

If estimates on exploration costs and volume estimates are available, the calibration of the model is relatively simple. We demonstrate how the model can be used to assess whether exploration costs warrant the potential benefits from finding reserves and extracting them. Specifically, we show how to calculate the value of the option to delay investment and discuss the agent's optimal investment threshold. This threshold, also referred to as the exercise boundary, depends on a number of variables and factors including: the agent's estimate of the volume in the reserve, the rate at which the agent learns about the volume of the reserve, the rate at which the agent extracts the commodity, and the expiry of the option. We show that the value of the option to wait-and-learn is high at the beginning and gradually decreases as expiry of the option approaches because the agent has little time left to learn.

We assume that the investment cost depends on the volume of the reserve. If the estimated volume is high (resp. low) the sunk cost to extract the commodity is high (resp. low). This has an effect on the optimal time to make the investment as well as the level of spot price commodity that justifies making the sunk cost. For example, we show that  when the volume estimate is low, and the option to invest is  far away from expiry, the agent sets a high investment threshold as a result of two effects which make the option to delay investment valuable. First,  low volume requires a high commodity spot price to justify the investment. Second, far away from expiry the investor attaches high value to waiting and learning about the volume estimates of the reserve. On the other hand, as the option approaches expiry, these two effects become weaker. In particular,  the value of learning is low because there is less time to learn about the reserve estimates, so  the investment decision is merely based on whether  costs will be recovered given the spot price of the commodity, the rate at which the commodity is extracted, and the uncertainty around it.

The remainder of this paper is organized as follows. In Section \ref{sec:model assumptions}, we provide the details of our modeling framework, including how we model both technical uncertainty and the uncertainty in the underlying project. Moreover, we provide an approach for accounting for the agent's learning of the reserve environment through exploration. Next, Section \ref{sec:real option valuation} provides an analysis of the Fourier space-time stepping approach for valuing the early exercise features in the irreversible investment with learning. Section \ref{sec:calibration} shows how the model can be calibrated to estimates of the cost of exploration and the expected benefits of this exploration. Section \ref{sec:results} provides some numerical experiments to demonstrate the efficacy of the approach and an analysis of the qualitative behavior of the model and its implications. Finally, we provide some concluding remarks and ideas for future lines of work.

\section{Model Assumptions} \label{sec:model assumptions}

In this section we provide models for the two sources of uncertainty  that drive the value of the real   option to explore and (irreversibly) invest in a project. The setting described here is tuned to some extent for oil exploration, however, it can be modified to deal with other activities  including: mining, natural and shale gas,   and other natural explorations.  Another extension to our setup is to account for the option to mothball exploration  and/or extraction (once extraction begins), as well as other managerial flexibilities that might arise in exploration and investment.  See for example \cite{McDonaldSiegel85}, \cite{DixitPindyck94}, \cite{Trigeorgis1999}, \cite{tsekrekos2010effect}, \cite{jaimungal2015incorporating}, and \cite{kobari2014real}.
We first describe how we model technical uncertainty and then describe how we model project value uncertainty. As usual, we assume a filtered probability space $\left( \Omega,\Ft,\{\Ft_t\}_{t \geq 0},\PP \right)$ where the filtration $\{\Ft_t\}_{t \geq 0}$ will be described in more detail at a later stage. We also assume the existence of an equivalent martingale measure, or risk-neutral measure, $\mathbb{Q}$ and all models below are written in terms of this probability measure.

\subsection{Technical Uncertainty}

Let $V=(V_t)_{t\ge0}$ denote the estimated reserve volume (level) process and $\hV$ be the true reserve volume. The true reserve volume $\hV$ is unknown to the investor, and can be viewed as a random variable when conditioned\footnote{For which we need to assume that $\hV < \infty$ almost surely.} on the information available to her at time $t$, but it will be revealed as $t\to\infty$. We assume for simplicity that the possible reserve levels, as well as their estimates, take on values from a finite set of possible reserve volumes. We  model the estimated reserve level as a continuous-time (inhomogeneous) Markov chain because reserve estimates are updated as new information from exploration is obtained. Moreover, to capture the feature that the accuracy of estimates improves as more information becomes available through time, we   assume that the transition rate between the volume estimate states decreases as time flows forward. In addition, we assume that $\displaystyle\lim_{t\to\infty}V_t = \hV$ almost surely to reflect the feature mentioned earlier that the true reserve level is revealed to the investor with the passage of time.\footnote{We could in principle derive such a model by writing $V_t=\EE[\hV|\mathcal G_t]$ where $\mathcal G=(\mathcal G_t)_{t\ge0}$ denotes the filtration that generates our information about the true reserve. Under some specific modeling assumptions, $V$ can be cast into a Markov chain representation. We opt not to delve into these details, as it detracts from the simplicity of the approach we are proposing, and instead model $V$ directly.}

More specifically, the estimated  reserve volume $V_t$ is modulated by a finite state, continuous-time, inhomogeneous Markov chain $Z_t$ taking values in $\{1,\dots,m\}$ via
\begin{equation}
V_t = v^{(Z_t)}\,,
\end{equation}
where the constants
\begin{equation}\label{eqn: possible reserve volumes}
\left\{v^{(1)}, \cdots, v^{(m)}\right\}\in\mathds R^m_+
\end{equation}
are the possible reserve volumes. The  generator matrix of the Markov chain $Z_t$ is denoted by $\bG_t$  and assumed to be of the form
\begin{equation}\label{eqn: generator matrix}
\bG_t = h_t\,\bA\,,
\end{equation}
where $h_t$ is a deterministic, non-negative decreasing function of time, such that $h_t \xrightarrow[]{t\to\infty} 0$ and $\int_0^\infty h_u\,du<\infty$, and $\bA$ is a constant $m\times m$ matrix with $\sum_{j=1}^n A_{ij} = 0$ and $A_{ij}>0$ for $i\ne j$. The states of the Markov chain correspond to various possible estimates for reserve level, thus capturing the uncertainty in those estimates.


The function $h_t$ captures how the agent learns about the volume or quantity of the commodity in the reserve. A decreasing $h$ implies that the transition rates are also decreasing, and hence the probability of changes in  the estimated  volumes  decreases with time, and therefore the estimates become more accurate. Optimal policies for the irreversible investment to explore, and the subsequent value of the project based on the extraction of the commodity, naturally  depend on the observed estimate of reserves -- Section \ref{sec:calibration} discusses in detail the form of $h_t$ and $\bA$, and how they are calibrated to data.

\subsection{Market Uncertainty}

The second source of uncertainty stems from the spot price of the commodity which we denote by  $S=(S_t)_{t\ge0}$, and is modelled as the exponential of an Ornstein-Uhlenbeck (OU) process:
\begin{subequations}\label{eqn:S model}
\begin{equation}\label{eqn:S model S}
S_t = \exp\{\theta+X_t\}\,,
\end{equation}
where the OU process $X=(X_t)_{t\ge0}$ satisfies the stochastic differential equation (SDE)
\begin{equation}\label{eqn:S model X}
dX_t = -\kappa \, X_t\,dt + \sigma \,dW_t\,,
\end{equation}
\end{subequations}
where $W=(W_t)_{t\ge0}$ is a standard Brownian motion (independent of $Z$), $\kappa>0$ is the rate of mean-reversion, $\theta$ is the (log-)level of mean-reversion, and $\sigma$ is the (log-)volatility of the spot price. Such models of commodity spot prices have been widely used in the literature, see for example \cite{CarteaFigueroa2005}, \cite{weron2007modeling}, \cite{CARTEA2008829}, \cite{kiesel2009two}, \cite{Coulon2013976}.

Now that we have specified the model for the stock of the commodity in the reserve and its market price, we need one final ingredient: the market value of the commodity in the reserve. We denote this value  by the process $P=(P_t)_{t\ge0}$ and  show how to calculate it in steps.

Suppose that investment is made at time $t$,   which is followed by extraction of the commodity  $\epsilon\geq 0$ later,  and extraction continues until the random (stopping) time $\tau=t+\epsilon+\Delta$. Here $\Delta>0$ represents the amount of time required to complete the extraction process. The investor does not know how much of the commodity is in the reserve, so the time to depletion (and the extraction duration) is a random stopping time. We also note that engineering and physical limitations  prevent the total amount of the commodity stored in the reserve from being extracted, and instead only $\gamma\,\hV$ is extractable, $0<\gamma<1$.\footnote{A good example is stored natural gas where there is always a residual that cannot be extracted from storage.}

\begin{figure}[t!]
\begin{center}
\begin{minipage}{0.58\textwidth}
\includegraphics[width=\textwidth]{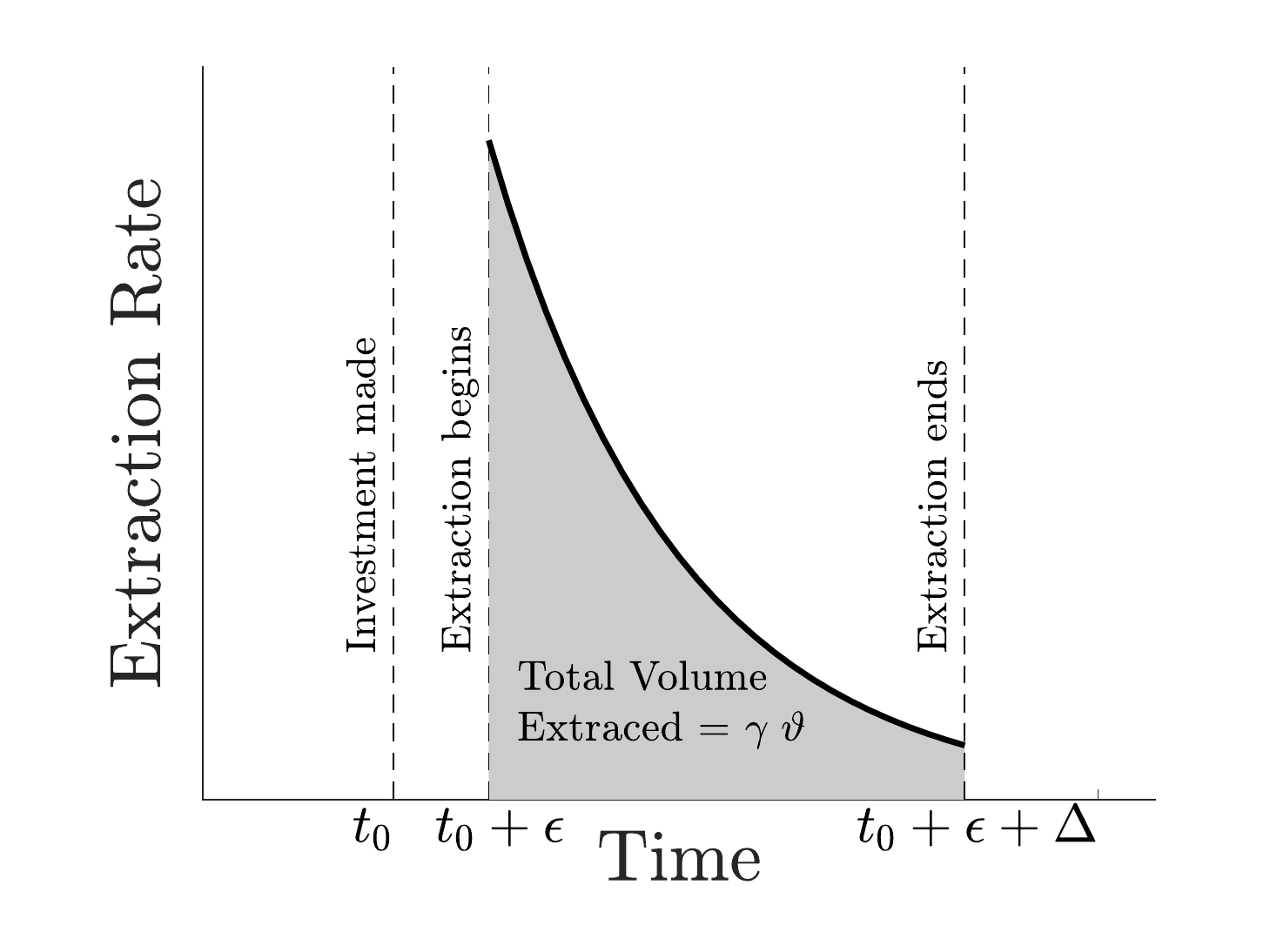}
\end{minipage}
\hfill
\begin{minipage}{0.4\textwidth}
\caption{Once the agent invests in the reserve at time $t$, extraction begins a time $t+\epsilon$ and continues until $\gamma$ percentage of the actual (unknown) reserve volume $\hV$ is extracted, which occurs at the random time $t+\epsilon+\Delta$.\label{fig:ExtractionRate}}
\vspace{3em}
\end{minipage}
\end{center}
\end{figure}

The random time to depletion $\Delta$ is related to the unknown total reserve volume $\hV$ via the rate of extraction. In particular, we have:
\begin{equation}
\int_{t+\epsilon}^{t+\epsilon+\Delta} g(u) \, du = \gamma\,\hV\,,
\end{equation}
where $g(u)$ denotes the rate of extraction at time $u$. We assume that once extraction begins the commodity is extracted at the rate
\begin{equation}\label{eqn:extraction rate}
g(u) = \alpha \,e^{-\beta(u-(t+\epsilon))}\,,\qquad u\in[t+\epsilon,t+\epsilon+\Delta]\,,
\end{equation}
where $\alpha\geq 0$, and $\beta\geq 0$.  Figure \ref{fig:ExtractionRate} presents a stylized picture of  the exponential extraction rate \eqref{eqn:extraction rate}. Under the specific extraction rate model in \eqref{eqn:extraction rate}, the time to depletion can be written as
\begin{equation}\label{eqn: time to extraction Delta}
\Delta =  -\tfrac{1}{\beta} \log\left(1 - \tfrac{\beta}{\alpha}\, \gamma \, \hV \right)\,,
\end{equation}
which is a random variable because $\hV$ is only known as $t\to\infty$. 

The value of the reserve when extraction begins is determined by a number of factors including: the price of the commodity, the state of the Markov chain linked to reserve uncertainty, and the random time to exhaustion of the reserve. Specifically, the discounted value of the cash-flow generated from extracting the commodity at the rate $g(u)$ and selling it at the spot price $S_u$ is given by
\begin{equation}\label{eqn: DCF}
DCF_t = \int_{t+\epsilon}^{t+\epsilon+\Delta} e^{-\rho(u-t)} \,(S_u - c) \, g(u) \, du\,,
\end{equation}
where $\rho$ is the agent's discount factor for the level of risk she bears with the project and $c$ is a running cost that the investor incurs as long as the extraction operation lasts.

To compute the market value of the commodity in the reserve, which we denote by $P_t$, we calculate the expected discounted value of the cash flows attained from selling the extracted commodity. Namely, we insert in \eqref{eqn: DCF}: the spot price of the commodity \eqref{eqn:S model}, the time to extraction completion given in \eqref{eqn: time to extraction Delta},   and the extraction rate  \eqref{eqn:extraction rate}.  Finally, we  take expectations of $DCF_t$ to obtain
\begin{align}
P_t = \EE \left[ DCF_t \,|\,\Ft_t\,\right] = \EE \left[ \left.\int_{t+\epsilon}^{t+\epsilon+\Delta} e^{-\rho(u-t)} \,(F_t(u) - c) \, g(u) \, du\,\right|\,\Ft_t\,\right]\,,
\label{eqn:ReserveValue}
\end{align}
where $\Ft_t = \sigma\left( (S_u, V_u)_{u \in [0,t]} \right)$ is the natural filtration generated by both $S$ and $V$ (or equivalently the Markov chain $Z$ introduced earlier), the expectation is taken under the risk-neutral measure $\mathbb{Q}$ and
\begin{eqnarray*}
F_t(u)&=&  \EE\left[S_u\,|\,\Ft_t\right]\\
&=&\exp\left\{ \theta + e^{-\kappa(u-t)}\,x + \frac{\sigma^2}{4\,\kappa}\left(1-e^{-2\kappa(u-t)}\right)\right\}
\end{eqnarray*}
is the forward price of the underlying asset.

Thus, the expectation in \eqref{eqn:ReserveValue} is over the random time to depletion  $\Delta$, which further depends on the reserve volume $\hV$. To compute this expectation we require the conditional distribution of the unknown reserve level $\hV$, that is we require $\PP\left(\hV = v^{(j)}\,|\,\Ft_t\right)$. By our assumption that $\hV=\displaystyle\lim_{t\to\infty}V_t$ a.s., this requires determining the conditional distribution at time $t$ of the limit of the underlying Markov chain. Specifically,
\[
\PP\left(\hV = v^{(j)}\,\middle|\,\Ft_t\right) = \PP\left(\lim_{t\to\infty} V_t = v^{(j)}\,\middle|\,\Ft_t\right)= \PP\left(\lim_{t\to\infty}  Z_t = j\middle|Z_t=i\right) = \left[e^{H_t \bA} \right]_{ij}\,,
\]
where $H_t = \int^{\infty}_t h_u \,du$, the notation $[\,\cdot\,]_{ij}$ denotes the $ij$ element of the matrix in the square brackets, and recall that the matrix $\bA$ is defined above in \eqref{eqn: generator matrix}. Therefore, the final expression for the expected discounted value of the reserve is
\begin{equation}\label{eqn: reserve value}
P_t := p^{(Z_t)}(t,X_t)=\sum_{j=1}^m \left[e^{H_t\,\bA}\right]_{Z_t,j}\,\int_{t+\epsilon}^{t+\epsilon-\tfrac{1}{\beta} \log{\left(1 - \tfrac{\beta}{\alpha}\, \gamma \, v^{(j)}  \right)}} e^{-\rho(u-t)} \,(F_t(u) - c) \, g(u) \, du\,.
\end{equation}
This expression has two sources of uncertainty, the first stems from the spot price of the commodity, through the OU process $X_t$, and the second from the estimate of the reserve volume, through the state of the Markov chain $Z_t$. With the model of extraction rate being exponentially decaying in time, see \eqref{eqn:extraction rate},   it is possible to write the integral appearing in the right-hand side of   equation \eqref{eqn: reserve value} in terms of special functions, however, such a rewrite does not add clarity so we opt to keep the integral as shown above.

\section{Real Option Valuation} \label{sec:real option valuation}

Now that we have a model for the value of the reserve, we focus on the cost required to exploit the reserve of the commodity and the value of the flexibility to decide when to make the investment. The cost of investing in the reserve is irreversible and denoted by $I^{(k)}$, where $k$ is the regime of the  reserve volume estimate. Here we assume that the cost $I^{(k)}$ is linked to the volume estimate because extracting a large reserve will likely  require a larger up-front investment than that required to extract the commodity from a small reserve. We assume that the investment cost is
\begin{equation}\label{eqn: linear cost}
 I^{(k)} = c_0 + c_1\,v^{(k)}\,,
\end{equation}
where $c_0\geq0$ is a fixed cost, $c_1\geq 0$, and  recall that $v^{(k)}$  are the possible reserve volumes, see \eqref{eqn: possible reserve volumes}. Notice that the investment cost depends on the reserve estimate. This reflects the fact that the investor will base the size of their extraction operation on their estimate of the reserve amount: the larger they believe the reserve to be, the bigger the required operation. The running cost $c$ ensures that the investor's running cost is still tied to the true reserve amount.

We denote the value of the   option  by $L_t=\ell^{(Z_t)}(t,X_t)$, where the collection of functions $\ell^{(1)}(t,x)$, $\dots$, $\ell^{(m)}(t,x)$ represents the value of the real option conditional on the state $Z_t=1,\dots,m$ (indexed by the superscript) and $X_t=x$. The agent must make a decision by time $T$, or they will lose the option to make the investment and exploit the reserve.  Standard theory implies that the value of the option to irreversibly invest in the reserve, and begin extraction, is given by the optimal stopping problem:
\begin{subequations}
\begin{align}
L_t &=
\sup_{\tau\in\mathcal T}\EE\left[\,e^{-\rho \,\tau}\,\max\left(P_\tau - I^{(Z_\tau)},0\right) \,|\,\Ft_t\,\right] \\
&=\sup_{\tau\in\mathcal T}\EE\left[\,e^{-\rho \,\tau}\,\max\left(P_\tau - I^{(Z_\tau)},0\right) \,|\, Z_t,\, X_t \,\right] \,.
\end{align}
\end{subequations}
Here, $\mathcal T$ denotes the set of admissible stopping times,  taken to be the finite collection of $\Ft$-stopping times restricted to $t_i =i\Delta t$, $i=0,\dots,N$ with $t_N\le T$. In other words, the agent is restricted to making the investment decision on days $t_i$. In the interim time, the agent can acquire more information to improve her volume reserve estimates.

For notational convenience we define the deflated value process
\begin{equation}\label{eqn: deflated value process}
  \bar\ell^{(j)}(t,x):=e^{-\rho t}\ell^{(j)}(t,x)\,,
\end{equation}
and observe that in between the investment dates, the deflated value processes
$\bar\ell^{(j)}(t,x)$ for $j=1,\cdots,m$,
are martingales. In addition,  since in between the investment dates there is no opportunity to exercise the option, $\bar\ell^{(j)}(t,x)$  is the same as a European claim with payoff equal to the value at the next exercise date. Thus,
\begin{equation}\label{eqn: maximization}
\bar\ell^{(j)}(t_i,x) = \max\left(\;\lim_{t\downarrow t_i}\bar\ell^{(j)}(t,x) \, ;\, e^{-\rho\,t_i}\,\left(p^{(j)}(t,x) - I^{(j)}\;\right) \right)\,,
\end{equation}
where $p^{(j)}(t,x)$ is as in \eqref{eqn: reserve value}, and recall that $j = 1,\dots, m$ represents the state of the  regime.

Finally, in the interval $t\in(t_i,t_{i+1}]$ the   processes $\bar\ell^{(j)}(t,x) $ satisfy the coupled system of PDEs
\begin{equation}
(\partial_t+\mathscr{L})\,\bar\ell^{(j)}(t,x) + h_t \sum_{j=1}^m A_{jk}\, \bar\ell^{(k)}(t,x) = 0\,,\qquad t\in (t_i,t_{i+1}]\,, \label{eqn:RealOption_PDE}
\end{equation}
where $\mathscr{L}=-\kappa\,x\,\partial_x +\frac{1}{2}\sigma^2\partial_{xx}$ is the infinitesimal generator of the process $X_t$.

The maximization in \eqref{eqn: maximization} represents the agent's option to  hold on to the investment option at time $t_i$ or to invest immediately. If the second argument attains the maximum, then the agent exercises her option to invest in the reserve, at a cost of $I^{(j)}$, and receives the expected discounted value of the cash-flow $p^{(j)}(t,x)$, which results from extracting and selling the commodity on the spot market. This investment decision is tied to the reserve volume estimate through the regime $j$. Different regimes $j$ will result in different exercise policies and we explore this relationship in the next section.


Motivated by the work of \cite{Surkov11}, who study options on multiple commodities driven by L\'evy processes, we solve the system of PDEs \eqref{eqn:RealOption_PDE} recursively by employing the Fourier transform of $\bar\ell^{(j)}(t,x)$ with respect to $x$, which we denote by $\tilde{\ell}^{(j)}(t,\omega)$. Specifically, we write
\begin{equation}
\tilde{\ell}^{(j)}(t,\omega) = \int_{-\infty}^\infty e^{-\imath\,\omega\,x}\,\bar{\ell}^{(j)}(t,x)\,dx\,,
\quad \text{and} \quad
\bar{\ell}^{(j)}(t,x) = \int_{-\infty}^\infty e^{\imath\,\omega\,x}\,\tilde{\ell}^{(j)}(t,\omega)\,\frac{d\omega}{2\pi}\,,
\end{equation}
where $\imath=\sqrt{-1}$.
Applying the Fourier transform to \eqref{eqn:RealOption_PDE}, we obtain a new PDE, without the parabolic term, which depends on the  state variable $\omega$ rather than the  state variable $x$, i.e.:
\begin{equation}
\left[ \partial_t + (\kappa-\tfrac{1}{2}\sigma^2\,\omega^2)+\kappa\,\omega\,\partial_\omega \right] \tilde{\ell}^{(j)}(t,\omega) + h_t \sum_{j=1}^m A_{jk}\,  \tilde{\ell}^{(k)}(t,\omega)=0\,.
\end{equation}
Within the interval $(t_{k},t_{k+1}]$, we introduce a moving coordinate system and write $\hat\ell^{(j)}(t,\omega) = \tilde\ell^{(j)}(t,e^{-\kappa(t_{k+1}-t)}\omega)$, which removes the derivative in $\omega$ and we find that the functions $\hat\ell^{(j)}$ satisfy the coupled linear system of ODEs
\begin{equation}
\partial_t \hat\ell^{(j)}(t,\omega)+ \left(\kappa-\tfrac{1}{2}\sigma^2\,\omega^2\,e^{-2\kappa(t_{k+1}-t)}\right)\hat\ell^{(j)}(t,\omega)
+ h_t \sum_{j=1}^m A_{jk}\,  \hat{\ell}^{(k)}(t,\omega) = 0\,.
\end{equation}

By writing $\bA=\boldsymbol{U}\boldsymbol{D}\boldsymbol{U}^{-1}$ where $\boldsymbol{U}$ is the matrix of eigenvectors of $\bA$, and $\boldsymbol{D}$ the diagonal matrix of eigenvalues of $\bA$,   the above coupled system of ODEs can be recast as independent ODEs, which in vector form reads
\begin{equation}
\partial_t \left(\boldsymbol{U}^{-1} \boldsymbol{\hat\ell}(t,\omega) \right)
+ \left(\psi(\omega\,e^{-\kappa(t_{k+1}-t)})\,\mathbb{I} +h_t\,\boldsymbol{D}\right)\boldsymbol{U}^{-1}\boldsymbol{\hat\ell}(t,\omega) = \boldsymbol{0}\,,
\end{equation}
where $\boldsymbol{\hat\ell}(t,\omega) = (\hat\ell^{(1)}(t,\omega),\dots,\hat\ell^{(n)}(t,\omega))'$, $\psi(\omega) = \kappa-\frac{1}{2}\sigma^2\,\omega^2$ and $\mathbb{I}$ is the $n\times n$ identity matrix. These uncoupled ODEs have solution
\begin{equation}\label{eqn: solution uncoupled ode}
\boldsymbol{U}^{-1}\boldsymbol{\hat\ell}(t_k^+,\omega) = \exp\left\{
\int_{t_k}^{t_{k+1} } \psi(\omega\,e^{-\kappa(t_{k+1}-s)})\,ds\, \mathbb{I} +
\int_{t_k}^{t_{k+1} } h_s\,ds\,\boldsymbol{D}\right\} \,\boldsymbol{U}^{-1}\boldsymbol{\hat\ell}(t_{k+1},\omega)\,.
\end{equation}
where $\boldsymbol{\hat\ell}(t_{k_+},\omega)= \displaystyle \lim_{t\downarrow t_k}\boldsymbol{\hat\ell}(t,\omega)$.

Next, we left-multiply by $\boldsymbol{U}$ to obtain
\begin{equation}
\boldsymbol{\hat\ell}(t_k^+,\omega) = \exp\left\{
\int_{t_k}^{t_{k+1} } \psi\left(\omega\,e^{-\kappa(t_{k+1}-s)}\right)\,ds\,\right\}
\exp\left\{ \int_{t_k}^{t_{k+1} } h_s\,ds\,\bA\right\} \, \boldsymbol{\hat\ell}(t_{k+1},\omega)\,,
\end{equation}
and  the Fourier transform of the deflated value of the option to irreversibly invest is
\begin{equation}
\boldsymbol{\tilde\ell}(t_k^+,\omega) = \exp\left\{
\int_{0}^{t_{k+1}-t_k } \psi(\omega\,e^{\kappa\,s})\,ds\,\right\}
\exp\left\{ \int_{t_k}^{t_{k+1} } h_s\,ds\,\bA\right\} \, \boldsymbol{\tilde\ell}\left(t_{k+1},\omega\,e^{\kappa(t_{k+1}-t_k)}\right)\,.
\end{equation}

This result has a few interesting features. The first is that the role of mean-reversion decouples from the Markov chain driving the volume estimates. The second is that the value at time $t^+_k$  at frequency $\omega$ depends on the value at time $t_{k+1}$ at frequency $\omega\,e^{\kappa(t_{k+1}-t_k)}$. This requires an extrapolation in the frequency space as the algorithm to calculate the option value steps backward in time. When we discretize the state space, such extrapolations could lead to inaccurate results since the edges of the state space are the most important contributions to the extrapolated values. Instead, we make use of the inverse relationship between frequencies and real space in Fourier transforms
\[
\int_{-\infty}^\infty e^{\imath\,\omega\,x}\,g(x/a)\,dx = \int_{-\infty}^\infty e^{\imath\,(a\,\omega)\,x}\,g(x)\,\frac{dx}{a} = \tfrac{1}{a}\,\tilde{g}(a\,\omega)\,,
\]
to write
\[
\boldsymbol{\tilde\ell}(t_k^+,\omega) = \exp\left\{
\int_{0}^{t_{k+1}-t_k } \psi(\omega\,e^{\kappa\,s})\,ds\,\right\}
\exp\left\{ \int_{t_k}^{t_{k+1} } h_s\,ds\,\bA\right\} \,
\boldsymbol{\tilde{\breve\ell}}(t_{k+1},\omega)\,,
\]
where $\boldsymbol{\breve\ell}(t_{k+1},x) = \boldsymbol\ell(t_{k+1},x\,e^{-\kappa(t_{k+1}-t_k)})$ and $\boldsymbol{\tilde{\breve\ell}}$ denotes the Fourier transform of $\boldsymbol{\breve\ell}$. Thus the value of the right-limit of the real option to invest at $t_k^+$ is determined in terms of interpolation in $x$ (rather than extrapolation in $\omega$). In Figure \ref{fig:algo} we  summarize the approach for valuing the real option to irreversibly invest in the reserve.
\begin{figure}[h!]
{\small
\begin{tabular}{r}
\hline\hline
\hspace{0.8\textwidth}
\end{tabular}
\vspace{-1em}
\begin{enumerate}
\item \texttt{Set the real and frequency space grids\\
 $x=[-\overline{x}:\Delta x:\overline{x}]$, $\breve{x}=e^{-\kappa\Delta t}\,x$ and $\omega = [0:\Delta \omega: \overline{\omega}]$.}

\item \texttt{Place terminal conditions: $\boldsymbol{\ell}(t_n,x) = e^{-\rho\,t_n}(\boldsymbol{P}(t_n,x)-\boldsymbol{I})_+ $.}
\item \texttt{Set $k=n$.}

\item \texttt{Step backwards from $t_{k+1}$ to $t_k$}:
\begin{enumerate}
\item \texttt{$\boldsymbol{\breve{\ell}}_{t_{k+1}} = \texttt{interp}(x,\,\boldsymbol{\ell}_{t_{k+1}},\, \breve{x}) $}
\item \texttt{$\boldsymbol{\tilde{\breve{\ell}}}_{t_{k+1}}(\omega) = \texttt{F}[ \boldsymbol{\breve{\ell}}_{t_{k+1}}(x) ]$}
\item \texttt{$\boldsymbol{\breve{\ell}}_{t_k^+}(x) = \texttt{F}^{-1}\left[ e^{\int_{0}^{t_{k+1}-t_k } \psi(\omega\,e^{\kappa\,s})\,ds}\,e^{\int_{t_k}^{t_{k+1}} h_s\,ds\,\bA}\,\boldsymbol{\tilde{\breve{\ell}}}_{t_{k+1}}(\omega) \right]$ }
\item
\texttt{$\boldsymbol{\tilde{\ell}} _{t_k} = \max\left(\boldsymbol{\tilde{\ell}}_{t_k^+}(x)\;; e^{-\rho\,t_k}(\boldsymbol{P}(t_k,x)-\boldsymbol{I})_+\right)$}
\end{enumerate}
\item \texttt{Set $k\to k-1$, if $k\ge0$ go to step 4.}
\end{enumerate}
\begin{tabular}{r}
\hline\hline
\hspace{0.8\textwidth}
\end{tabular}
}
\vspace{-1em}
\caption{Algorithm for computing the value of the option to irreversibly invest in the reserve. \label{fig:algo}}
\end{figure}

\section{Calibration and learning}\label{sec:calibration}

Armed with the model in Section \ref{sec:model assumptions}, and the valuation procedure developed in Section \ref{sec:real option valuation}, we discuss in detail the calibration procedure for the Markov chain parameters and the investor's learning function $h_t$ given by \eqref{eqn: generator matrix}. The general idea is to use the investor's prior information regarding the estimate to define the possible reserve estimates, represented by the states of the Markov chain, as well as the transition rates for the base generator matrix $\bA$. The next step is to calibrate the learning function using information about how much the reserve estimate variance can be reduced by some future date.

\subsection{Calibration of Markov chain parameters}

At time $t=0$  the agent has an estimate of the true reserve volume, denoted by $\mu$, and the volatility of the estimate, denoted by $\sigma_0$. Thus
\begin{equation}
\mu=\EE[\hV\,|\,\Ft_0] \qquad\text{and}\qquad
\sigma_0^2=\VV\left[\hV\,|\,\Ft_0\right]\,.
\label{eqn:initial reserve volume distribution}
\end{equation}
The first step in the calibration procedure is to select  the states of reserve volume conditional on the Markov chain state, i.e. to select $v^{(1)},\dots,v^{(m)}$. Since the $t=0$ estimate of the reserve volume represents an unbiased estimator of the reserves, the Markov chain should be symmetric around the initial estimate of the reserve volume. To ensure this symmetry, we assume that (i) the cardinality of the states of the Markov chain $Z_t$ is odd, i.e., $m=2\,L+1$ for some positive integer $L$; (ii) $v^{(L+1)} = \mu$ where $\mu$ is the initial estimate of the reserve volume, (iii) $v^{(k)}$ is increasing in $k$  and (iv)
\begin{equation}
v^{(L+1+i)} - \mu = \mu - v^{(L+1-i)}, \qquad \forall \; i=1,\dots,L\,.
\end{equation}
We further assume that the agent's estimate of the volume is normally distributed (this assumption can be modified to any distribution the agent considers to represent her prior knowledge):
\begin{equation}
\hV \,|\,{\Ft_0}\sim \mathcal N(\mu, \;\sigma_0^2)\,,
\end{equation}
Therefore, we choose $v^{(k)}$ to be equally spaced over $n$ standard deviations of the normal random variable support, i.e., we select
\begin{equation}
v^{(k)} = \mu-n\,\sigma_0 + (k-1) \,\frac{2\,n\,\sigma_0}{m-1}\,, \qquad \forall\;k=1,\dots,m\,
\end{equation}

Placing symmetry on the states of the reserve volume estimator is not sufficient to ensure symmetry in its distribution. We further require the symmetry in the base generator rate matrix $\bA$, and assume that
\begin{subequations}
\begin{align}
\bA_{1,1}^\blambda &= -\lambda_1\,,  &\bA_{1,2}^\blambda &= \lambda_1\,,
\\
\bA_{i,i-1}^\blambda &= \lambda_i\,, &\bA_{i,i}^\blambda &= -2 \lambda_i , &
\bA_{i,i+1}^\blambda &= \lambda_i\,, & i = 2 \,,\dots, L\,,
\\
\bA_{L+1,L}^\blambda &= \lambda_{L+1}\,, & \bA_{L+1,L+1}^\blambda &= -2 \lambda_{L+1} , &
\bA_{L+1,L+2}^\blambda &= \lambda_{L+1}\,,
\\
\bA_{m-i,m-i-1}^\blambda &= \lambda_{i}\,, & \bA_{m-i,m-i}^\blambda &= -2\lambda_{i}, &
\bA_{m-i,m-i+1}^\blambda &= \lambda_{i}\,, & i = 1 \,,\dots, L-1\,,
\\
\bA_{m,m-1}^\blambda &= \lambda_1\,, & \bA_{m,m}^\blambda &= -\lambda_1\,.
\end{align}
\end{subequations}
for some set of $\blambda = \{\lambda_1,\lambda_2,\dots,\lambda_{L+1}\}$, where  $\lambda_1,\lambda_2,\dots,\lambda_{L+1}>0$. Here $\lambda_1$, $\left\{ \lambda_2,\dots,\lambda_{L} \right\}$ and $\lambda_{L+1}$ determine the transition rates out of the edge states, interior states  and the midstate, respectively. 
The form of $\bA^\blambda$ ensures transitions only occur between neighboring states and ensures symmetry in the transition rates of states that are across the mean estimate. The parameters $\blambda$ are calibrated so the invariant distribution of $\hV$ without learning  coincides with a discrete approximation of a normal random variable with mean $\mu$ and variance $\sigma_0^2$. This ensures that the Markov chain generates an invariant distribution equal to a discrete approximation of the original estimate of the reserve volume distribution. Formally, let $\bP^\blambda = e^{\bA^\blambda}$ denote the transition probability after one unit of time, and let $\bpi^\blambda$ denote the invariant distribution of $\bP$, i.e., $\bpi^\blambda$ solves the eigenproblem $\bP\,\bpi^\blambda=\bpi^\blambda$. Then, we choose $\blambda$ such that
\begin{subequations}
\begin{align}
\bpi_1^\blambda &= \Phi_{\mu,\sigma_0}\left( \tfrac{1}{2} \left( v^{(1)} + v^{(2)} \right) \right)    - \Phi_{\mu,\sigma_0}\left( \tfrac{1}{2} \left( v^{(1)} + \left[ v^{(1)} - \left(v^{(2)} - v^{(1)} \right) \right] \right)  \right)\,, \\
\bpi_2^\blambda &= \Phi_{\mu,\sigma_0}\left( \tfrac{1}{2} (v^{(i+1)} + v^{(i)}) \right)-
\Phi_{\mu,\sigma_0} \left( \tfrac{1}{2} (v^{(i)} + v^{(i-1)}) \right)\,, \quad i = 2,\dots,2L\,, \\
\bpi_m^\blambda &= \Phi_{\mu,\sigma_0}\left( \tfrac{1}{2} \left( v^{(m)} + \left[ v^{(m)} + \left(v^{(m)} - v^{(m-1)} \right) \right] \right)  \right) -
\Phi_{\mu,\sigma_0} \left( \tfrac{1}{2} (v^{(m)} + v^{(m-1)}) \right) \,,
\end{align}
\end{subequations}
where $\Phi_{\mu,\sigma_0}(\cdot)$ denotes the normal cumulative density of a normal with mean $\mu$ and variance $\sigma_0^2$. Note that $\Phi_{\mu,\sigma_0}$ is evaluated at the midpoint between possible reserve amounts and that we extrapolate linearly to obtain the equations for the edge states.

\subsection{Calibration of the learning function}

As time passes, the agent gathers more and better quality information about the volumes of the commodity in the reserve. Thus,  the variance of the estimated volume in the reserve is expected to decrease from $\sigma_0^2$  to $\sigma_{T'}^2 < \sigma_0^2$ by some  fixed time $T' < T$. Specifically\footnote{In principle, we could develop a model that can calibrate to a sequence of times and variances, however, the one step reduction is enough to illustrate the essential ideas in this reduction.}
\begin{equation}
\sigma_{T'}^2=\VV\left[\hV\,|\,\Ft_{T'}\right].
\label{eqn:learned reserve volume distribution}
\end{equation}
This parameter, along with the starting reserve estimate variance, $\sigma_0^2$, are the main determinants of the learning function. For parsimony, we assume that the agent's learning function is of the form
\[
 h_t = a\,e^{-b\,t} \quad \text{for some }a,b>0\,,
\]
where the parameter $a$ represents the initial transition rates, i.e. $h_0=a$, between the states of the Markov chain, and hence reflects the  uncertainty in the initial estimates of reserves.
The learning parameter $b$ represents the rate at which the agent learns -- the larger (resp. smaller) is $b$ the quicker is the learning process because large (resp. small) values of $b$  make  the transition rates decay  faster (slower) through time and therefore reserve estimates become stable quickly (resp. slowly). Recall that the learning rate function plays a key role in the generator matrix of the Markov chain, see \eqref{eqn: generator matrix}, in that it  captures how the agent learns how much volume of the commodity is in the reserve.


The parameters in the learning rate function $h$ are calibrated to obey the constraints \eqref{eqn:initial reserve volume distribution} and \eqref{eqn:learned reserve volume distribution}. Due to the symmetry in the base transition rates $\bA$, and the symmetry in the reserve volume states $v^{(k)}$, we automatically satisfy the mean constraints
\begin{equation}
\EE[\hV \;|\; V_0 = \mu] = \mu\,, \qquad \text{and} \qquad \EE[\hV\;|\; V_{T'} = \mu] = \mu\,.
\end{equation}
To satisfy the variance constraints we require the transition probabilities of the Markov chain from an arbitrary state at time $t$ to its infinite horizon state, which we denote $p_{t,ij} := \PP \left( \displaystyle \lim_{t\to\infty}  Z_t =j\,|\,Z_{t} = i \right)$, to be
\begin{equation}
p_{t,ij}  = \left[\exp\left\{\left({\textstyle\int_{t}^{\infty}} h_u\,du\right) \bA\right\} \right]_{ij} =
\left[\exp\left\{\tfrac{a}{b}\,e^{-b\,t}\bA\right\} \right]_{ij}\,.
\end{equation}

Note that the right-hand side of the equation above  is a matrix exponential and, as before, the notation $[\,\cdot\,]_{ij}$ denotes the $ij$ element of the matrix in the square brackets. Now we must solve the two coupled system of non-linear equations for the parameters $a$ and $b$:
\begin{subequations}
\begin{align}
\sigma_0^2 = \VV\left[\hV\,|\, V_0 = \mu\right] & = \sum_{k=1}^{2L+1} v^{(k)} \left( p_{0,Lk} - \mu \right) ^2\,, \label{eqn:initial variance matching}\\
\sigma_{T'}^2= \VV\left[\hV\,|\, V_{T'} = \mu\right] & = \sum_{k=1}^{2L+1} v^{(k)} \left( p_{T',Lk} - \mu \right) ^2 \,,
\end{align}%
\end{subequations}%
where $\VV[\cdot\,|\,\cdot]$ is the variance operator of the first argument conditioned on the event in the second argument.

After which, all technical uncertainty model parameters are calibrated to the distributional properties of the initial reserve volume estimates and the reduction in variance as a result of learning.

The value of the irreversible option to invest in the reserve with learning can now be valued using the approach in Section \ref{sec:real option valuation}, a summary of which is presented in the algorithm shown in Figure \ref{fig:algo}. The value of exploration, which improves the variance of the estimators of the reserve volume, can be assessed by considering the option value when there is no learning and comparing it to the option value without learning.

\section{Numerical Results}\label{sec:results}

In this section we investigate   the optimal exercise policies of the agent and assess the value of learning. Throughout we use the following  parameters and modeling choices:
\begin{itemize}
\item \textbf{Reserve volume}. Initial reserve estimate: $\mu = 10^9$, initial reserve estimate variance: $\sigma_0^2 = 3 \times 10^8$.
\item \textbf{Investment costs}. Fixed cost parameter: $c_0 = 10^8$, variable cost parameter: $c_1 = 3 \times 10^6$. This implies an investment cost of $1.12 \times 10^8$ when the reserve estimate is lowest and $1.48 \times 10^8$ when the reserve estimate is highest.
\item \textbf{Expiry of option}. $T=5$ years which consists of 255 weeks.
\item \textbf{Other model parameters:}
\begin{itemize}

\item Underlying resource model parameters: $\kappa = 0.5$, $\theta = \log(100)$, $\sigma_X = 0.5$.

\item Discount rate: $\rho = 0.05$.

\item Extraction rate parameters: $\alpha = 1$, $\beta = 0.05$, $\gamma = 0.9$, $\epsilon = 2$.

\item Markov chain parameters: 31 states.
\end{itemize}

\end{itemize}

Furthermore, we will consider the case of slow learning and fast learning which we model by changing the reduction in reserve estimate variance ($\sigma_{T'}^2 = 2.5 \times 10^8$ versus $\sigma_{T'}^2 = 10^8$) with $T' = 2$ years. We also will consider the case with and without running costs ($c=20$ and $c=0$, respectively). Finally, we compare these to the case where this is no learning by setting $a = 1$ and $b = 0$.

\begin{figure}[t!]
\centering
\includegraphics[width=\textwidth]{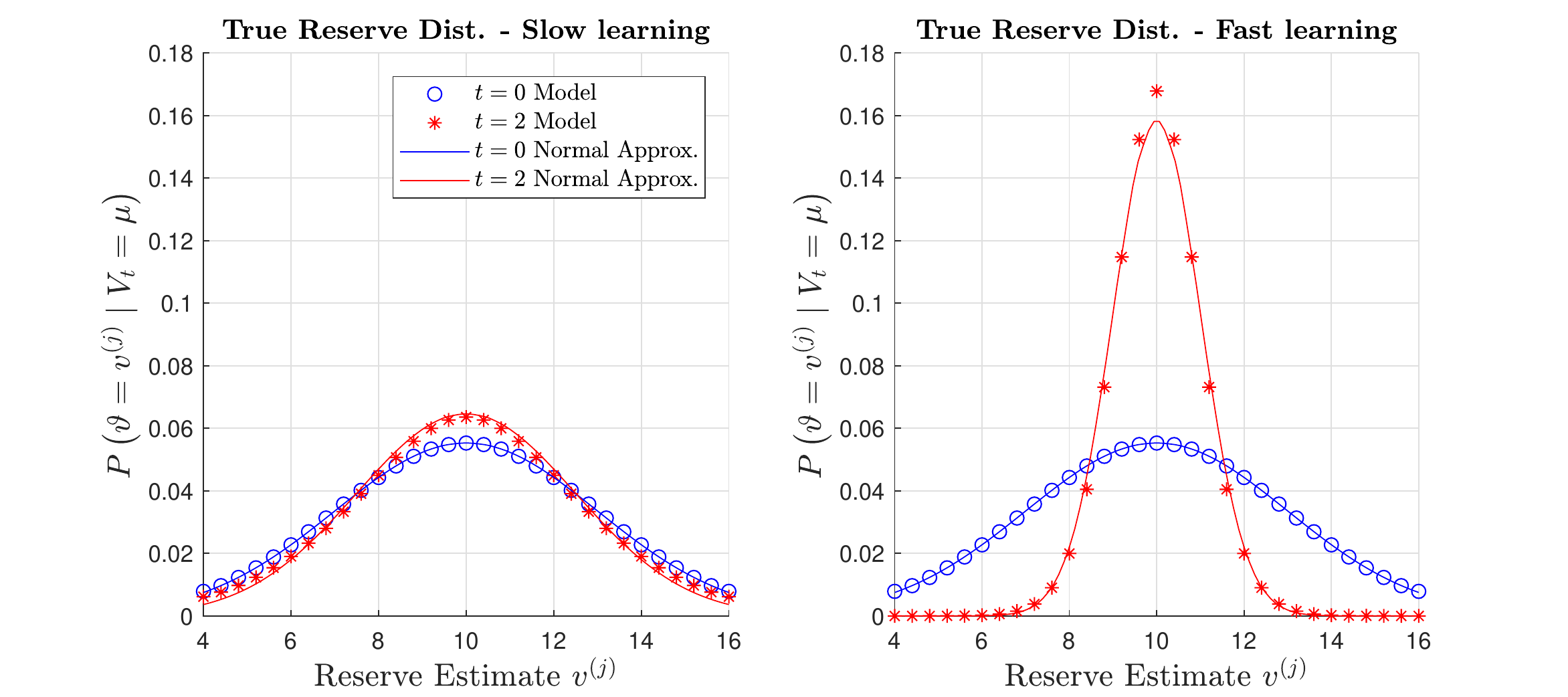}
\caption{Model-implied distribution of true reserve amount at $t=0$ and $t=2$ (variance reduction timeframe) conditional on being at the initial estimate $\mu = v^{(L+1)}$ in the case of slow learning (left panel) and fast learning (right panel). Lines correspond to normal approximations for the invariant distribution with parameters $\left( \mu, \sigma^2_0 \right)$ and $\left(\mu, \sigma^2_{T'} \right)$, respectively.}
\label{fig:calibration}
\end{figure}




Figure \ref{fig:calibration} shows the effect of learning on the distribution of the true reserve amount. We find that in both the slow and fast learning cases, the distribution (conditional on being at the initial reserve estimate) is more peaked around the mid-state. This reflects the fact that the investor has learned more about the reserve amount and, given that they are in the mid-state at $t=2$ (the timeframe given for calibrated variance reduction), they are now more inclined to believe that $\mu$ is the true reserve amount. This effect is slight in the slow learning case and more pronounced with fast learning as the investor is more confident having learned more about the reserve in the same amount of time. Note that the model-implied probabilities match the normal approximation only at $t=0$ since this was the only constraint imposed in the calibration procedure.

\begin{figure}[h!]
\centering
\includegraphics[width=\textwidth]{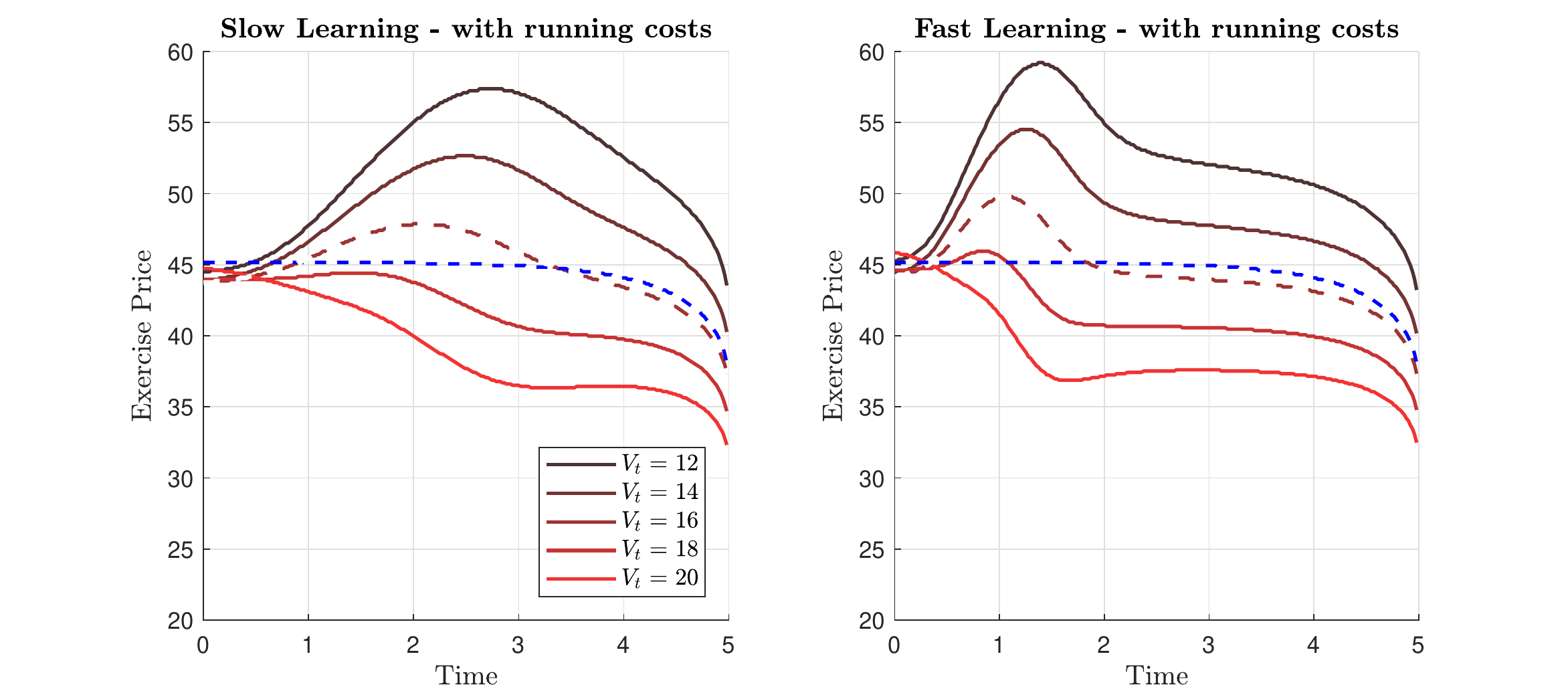}
\caption{Exercise boundary for different reserve estimates states when the agent's learning rate is low (left panel) vs. high (right panel) when they incur running costs. Brighter lines correspond to higher reserve estimates, the dashed red line corresponds to the middle state (initial reserve estimate) and the blue line is the exercise boundary when there is no learning.}\label{fig:exBoundary_withCosts}
\end{figure}

Figure \ref{fig:exBoundary_withCosts} shows the optimal exercise boundary for an agent who learns at different rates and for different volume estimates (i.e. different states of the Markov chain). For simplicity, we only display a select number of states near the mid-state as these represent the most relevant volume estimates for the investor. The $y$-axis of the figure shows the spot price of the commodity at which the investor would exercise the option, and the $x$-axis is the time elapsed measured in years.  The figure shows that as the agent's volume estimate increases, the exercise boundary shifts down because a larger reserve requires a lower commodity spot price to justify the investment.

We observe that the inclusion of learning, both fast and slow, has a profound effect on the shape of the exercise boundary. In the no-learning case, this boundary is non-increasing with time whereas the cases of slow and fast learning lead to exercise boundaries with non-trivial shapes that include increases and decreases at different rates, with behavior varying depending on the investor's prevailing reserve estimate. The explanation for the general shape these curve takes is as follows: when the investor is in a low state (i.e. has a low reserve estimate) the boundary is increasing to allow the learning process to potentially drive the estimate upward leading to a higher project value in the event that the reserve estimate increases. Eventually, the investor believes that the learning process has provided enough information to be confident that whatever state they currently occupy is in fact the true reserve amount. At this point the exercise boundary goes back to the typical decreasing pattern. The time at which this ``inflection point'' occurs depends on the rate of learning: when learning is fast, this turnaround comes sooner.

\begin{figure}[h!]
\includegraphics[width=\textwidth]{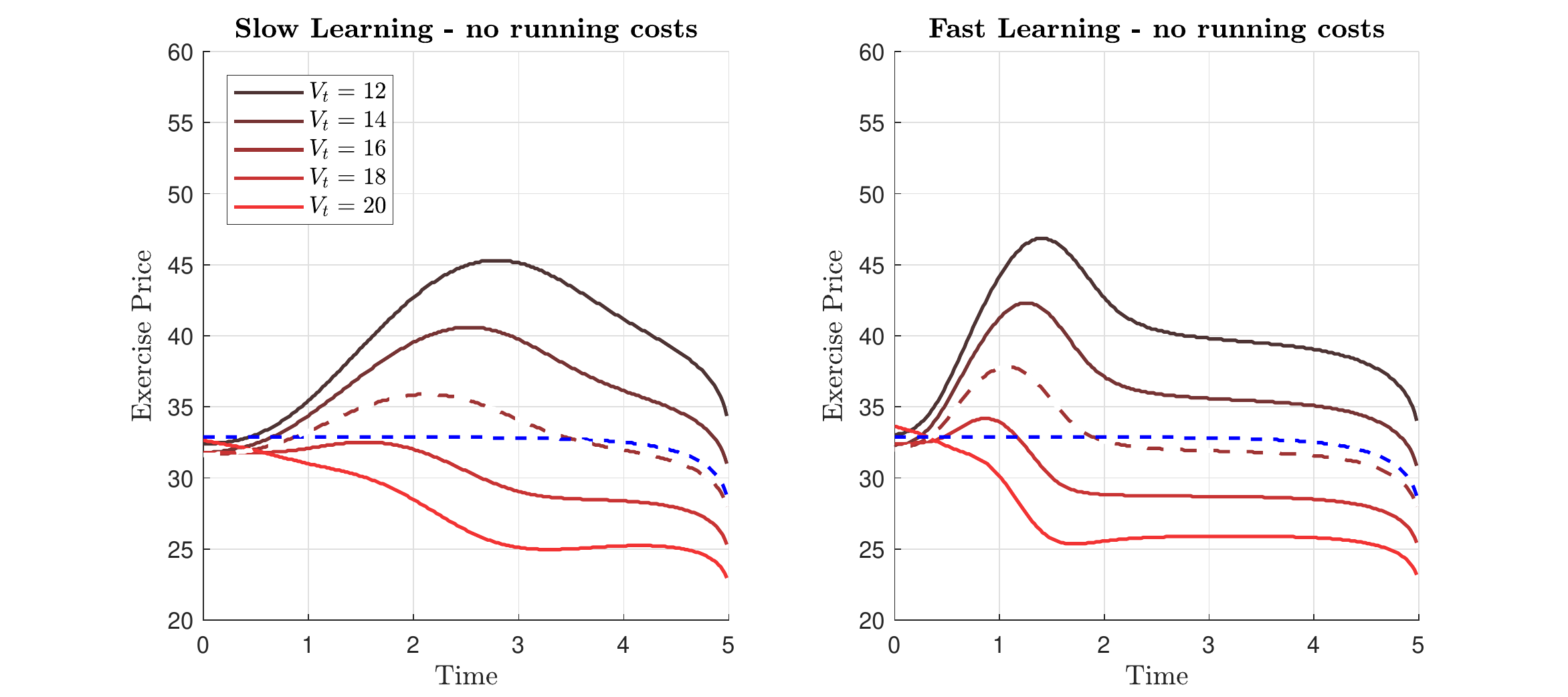}
\caption{Exercise boundary for different reserve estimates states when the agent's learning rate is low (left panel) vs. high (right panel) when they do not incur running costs. Brighter lines correspond to higher reserve estimates, the dashed red line corresponds to the middle state (initial reserve estimate) and the blue line is the exercise boundary when there is no learning.}\label{fig:exBoundary_noCosts}
\end{figure}

Figure \ref{fig:exBoundary_noCosts} shows the exercise boundaries, for different volume states, when the agent does not incur running costs. Observe that the exercise boundaries in both the fast and slow learning case have the same general shape as the boundaries in Figure \ref{fig:exBoundary_withCosts}. The effect of adding running costs is an upward parallel shift of the exercise boundaries, as the commodity spot price must be higher to justify the investment after accounting of running costs.

\begin{figure}[h!]
\centering
\includegraphics[width=\textwidth]{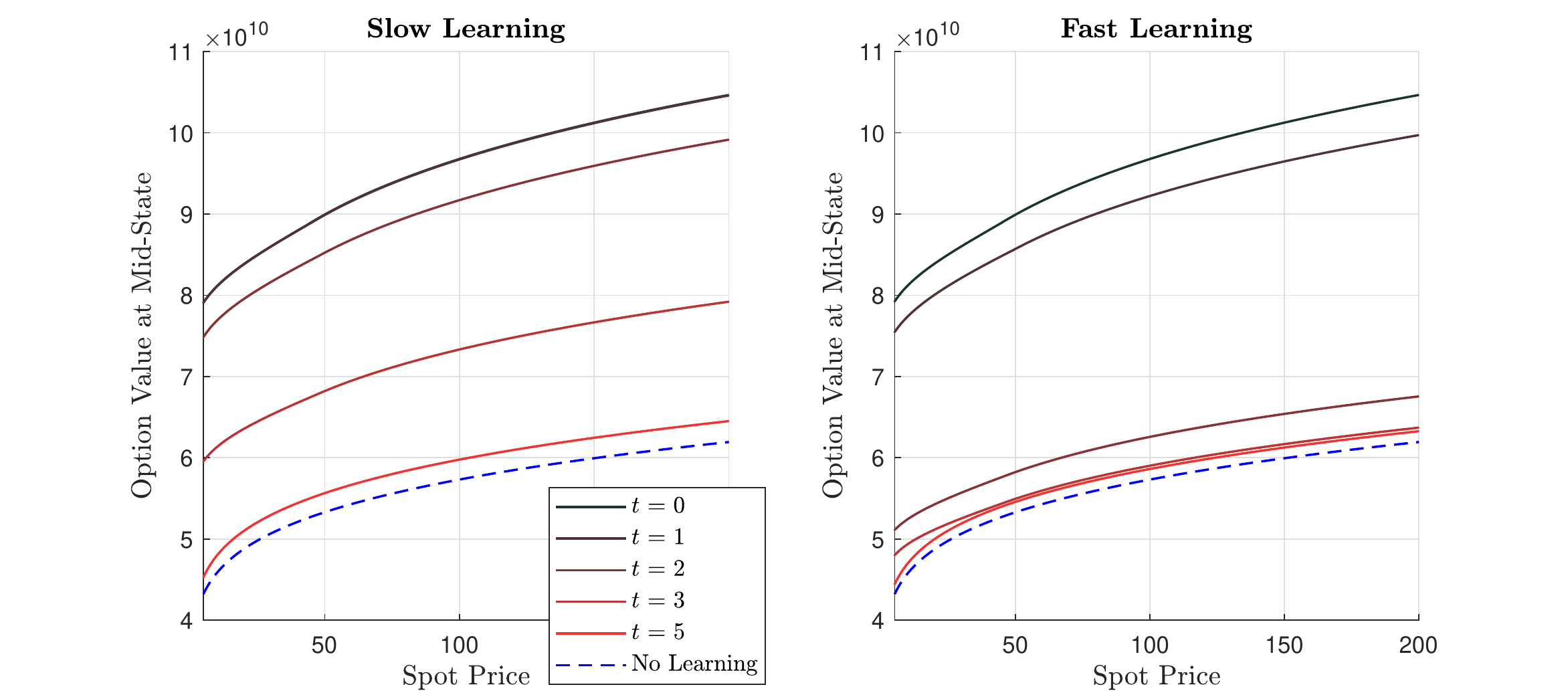}
\caption{Option value as a function of spot price through time assuming the reserve estimate is equal to the mid-state (initial estimate), $V_t = \mu$, with slow learning (left panel) vs. fast learning (right panel). Dashed lines correspond to option value at maturity when the investor does not learn.}\label{fig:optionValueThruTime}
\end{figure}

Figure \ref{fig:optionValueThruTime} compares the value of the option at different points in time for different learning rates, assuming that the investor's estimate is equal to the initial estimate. We compare this to the case of no learning, which is very insensitive to time-to-maturity relative to the learning case.\footnote{Note that since the value of the option changes very slightly with time-to-maturity in the no-learning case, we plot the no-learning option value curve only at $t=0$ in Figure \ref{fig:optionValueThruTime}.} We find that the value of the option to wait-and-learn is high at the beginning and gradually decreases as expiry of the option approaches because the agent has little time left to learn. As expected, this decline in value is accelerated in the fast learning case as the investor becomes more confident about the reserve amount at an earlier point in time. Note that the effect of the passage of time on the option value is different in other states.

\section{Conclusions}

In this paper we show how to incorporate technical uncertainty into the decision to invest in a commodity reserve. This uncertainty stems from not knowing the volume of the commodity stored in the reserve, and compounds with the uncertainty of the value of the reserve because future spot prices are unknown.

The agent has the option to wait-and-see before making the irreversible investment to exploit the commodity reserve. In our model, as time goes by,  the agent  learns about the volume of the commodity stored in the reserve, so the option to delay investment is valuable because it allows the agent to learn and to wait for the optimal market conditions (i.e. spot price of the commodity) before sinking the investment.

We adopt a continuous-time Markov chain to model the reserve volume and the technical uncertainty. In our model  the agent learns about the volume in the reserve as time goes by and the accuracy of the estimates is also improved with time. We show how to calculate the value of the option to delay investment and discuss the agent's optimal investment threshold.

We show how the exercise boundary depends on the agent's estimate of the volume (which depends on the Markov chain state) and how this boundary depends on the rate at which she refines her estimates of the reserve. For example, we show that when the option to invest is far away from expiry, and the volume estimate is low, the value attached to waiting and gathering more information is higher for the agent who can quickly learn about the volume of the reserves than for an agent who learns at a very low rate.

\newpage

\section{References}

\bibliography{TechnicalUncertainty}

\begin{thebibliography}{}

\bibitem[\protect\citeauthoryear{Armstrong, Galli, Bailey, and
  Cou\"{e}t}{Armstrong et~al.}{2004}]{ArmstrongGalliBaileyCouet}
Armstrong, M., A.~Galli, W.~Bailey, and B.~Cou\"{e}t (2004).
\newblock Incorporating technical uncertainty in real option valuation of oil
  projects.
\newblock {\em Journal of Petroleum Science and Engineering\/}~{\em 44\/}(1),
  67--82.

\bibitem[\protect\citeauthoryear{Brennan and Schwartz}{Brennan and
  Schwartz}{1985}]{BrennanSchwartz1985}
Brennan, M.~J. and E.~S. Schwartz (1985).
\newblock Evaluating natural resource investments.
\newblock {\em Journal of Business\/}~{\em 58\/}(2), 135--157.

\bibitem[\protect\citeauthoryear{Cartea, Donnelly, and Jaimungal}{Cartea
  et~al.}{2017}]{CDJ}
Cartea, {\'A}., R.~F. Donnelly, and S.~Jaimungal (2017).
\newblock Algorithmic trading with model uncertainty.
\newblock {\em SIAM Journal on Financial Mathematics\/}~{\em 8\/}(1), 635--671.

\bibitem[\protect\citeauthoryear{Cartea, Donnelly, and Jaimungal}{Cartea
  et~al.}{2018}]{Dempster}
Cartea, {\'A}., R.~F. Donnelly, and S.~Jaimungal (2018, march).
\newblock {\em High-Performance Computing in Finance: Problems, Methods, and
  Solutions\/} (1st ed.)., Chapter Portfolio liquidation and ambiguity
  aversion, pp.\ ~x.
\newblock CRC Financial Mathematics Series. Chapman and Hall.

\bibitem[\protect\citeauthoryear{Cartea and Figueroa}{Cartea and
  Figueroa}{2005}]{CarteaFigueroa2005}
Cartea, {\'A}. and M.~G. Figueroa (2005).
\newblock Pricing in electricity markets: a mean reverting jump diffusion model
  with seasonality.
\newblock {\em Applied Mathematical Finance\/}~{\em 12(4)}, 313--335.

\bibitem[\protect\citeauthoryear{Cartea and Gonz{\'a}lez-Pedraz}{Cartea and
  Gonz{\'a}lez-Pedraz}{2012}]{cartea2012much}
Cartea, {\'A}. and C.~Gonz{\'a}lez-Pedraz (2012).
\newblock How much should we pay for interconnecting electricity markets? {A}
  real options approach.
\newblock {\em Energy Economics\/}~{\em 34\/}(1), 14--30.

\bibitem[\protect\citeauthoryear{Cartea and Jaimungal}{Cartea and
  Jaimungal}{2017}]{cartea2017irreversible}
Cartea, {\'A}. and S.~Jaimungal (2017).
\newblock Irreversible investments and ambiguity aversion.
\newblock {\em International Journal of Theoretical and Applied Finance\/}~{\em
  20\/}(07), 1750044.

\bibitem[\protect\citeauthoryear{Cartea, Jaimungal, and Zhen}{Cartea
  et~al.}{2016}]{CJZ}
Cartea, {\'A}., S.~Jaimungal, and Q.~Zhen (2016).
\newblock Model uncertainty in commodity markets.
\newblock {\em SIAM Journal on Financial Mathematics\/}~{\em 7\/}(1), 1--33.

\bibitem[\protect\citeauthoryear{Cartea and Williams}{Cartea and
  Williams}{2008}]{CARTEA2008829}
Cartea, {\'A}. and T.~Williams (2008).
\newblock {UK} gas markets: The market price of risk and applications to
  multiple interruptible supply contracts.
\newblock {\em Energy Economics\/}~{\em 30\/}(3), 829 -- 846.

\bibitem[\protect\citeauthoryear{Cort\'azar, Schwartz, and Casassus}{Cort\'azar
  et~al.}{2001}]{CortozarSchwartzCasassus2000}
Cort\'azar, G., E.~S. Schwartz, and J.~Casassus (2001).
\newblock Optimal exploration investments under price and geological-technical
  uncertainty: a real options model.
\newblock {\em R\&D Management\/}~{\em 31\/}(2), 181--189.

\bibitem[\protect\citeauthoryear{Coulon, Powell, and Sircar}{Coulon
  et~al.}{2013}]{Coulon2013976}
Coulon, M., W.~B. Powell, and R.~Sircar (2013).
\newblock A model for hedging load and price risk in the {T}exas electricity
  market.
\newblock {\em Energy Economics\/}~{\em 40}, 976 -- 988.

\bibitem[\protect\citeauthoryear{Dixit}{Dixit}{1989}]{Dixit89}
Dixit, A. (1989).
\newblock Entry and exit decisions under uncertainty.
\newblock {\em The Journal of Political Economy\/}~{\em 97\/}(3), 620--638.

\bibitem[\protect\citeauthoryear{Dixit and Pindyck}{Dixit and
  Pindyck}{1994}]{DixitPindyck94}
Dixit, A. and R.~Pindyck (1994).
\newblock {\em Investment under Uncertainty}.
\newblock Princeton University Press.

\bibitem[\protect\citeauthoryear{Elliott, Miao, and Yu}{Elliott
  et~al.}{2009}]{ElliotMiaoYu2009}
Elliott, R.~J., H.~Miao, and J.~Yu (2009).
\newblock Investment timing under regime switching.
\newblock {\em International Journal of Theoretical and Applied Finance\/}~{\em
  12\/}(04), 443--463.

\bibitem[\protect\citeauthoryear{Fleten, Heggedal, and Siddiqui}{Fleten
  et~al.}{2011}]{fleten2011transmission}
Fleten, S.-E., A.~M. Heggedal, and A.~Siddiqui (2011).
\newblock Transmission capacity between {N}orway and {G}ermany: a real options
  analysis.
\newblock {\em The Journal of Energy Markets\/}~{\em 4\/}(1), 121.

\bibitem[\protect\citeauthoryear{Himpler and Madlener}{Himpler and
  Madlener}{2014}]{himpler2014optimal}
Himpler, S. and R.~Madlener (2014).
\newblock Optimal timing of wind farm repowering: a two-factor real options
  analysis.
\newblock {\em Journal of Energy Markets\/}~{\em 7\/}(3), 3.

\bibitem[\protect\citeauthoryear{Jackson, Jaimungal, and Surkov}{Jackson
  et~al.}{2008}]{Jai08}
Jackson, K., S.~Jaimungal, and V.~Surkov (2008).
\newblock Fourier space time stepping for option pricing with {L}evy models.
\newblock {\em Journal of Computational Finance\/}~{\em 12\/}(2), 1--29.

\bibitem[\protect\citeauthoryear{Jaimungal, De~Souza, and Zubelli}{Jaimungal
  et~al.}{2013}]{JaiSouzaZubelli09}
Jaimungal, S., M.~O. De~Souza, and J.~P. Zubelli (2013).
\newblock Real option pricing with mean-reverting investment and project value.
\newblock {\em The European Journal of Finance\/}~{\em 19\/}(7-8), 625--644.

\bibitem[\protect\citeauthoryear{Jaimungal and Lawryshyn}{Jaimungal and
  Lawryshyn}{2015}]{jaimungal2015incorporating}
Jaimungal, S. and Y.~Lawryshyn (2015).
\newblock Incorporating managerial information into real option valuation.
\newblock In {\em Commodities, Energy and Environmental Finance}, pp.\
  213--238. Springer.

\bibitem[\protect\citeauthoryear{Jaimungal and Surkov}{Jaimungal and
  Surkov}{2011}]{Surkov11}
Jaimungal, S. and V.~Surkov (2011).
\newblock L{\'e}vy-based cross-commodity models and derivative valuation.
\newblock {\em SIAM Journal on Financial Mathematics\/}~{\em 2\/}(1), 464--487.

\bibitem[\protect\citeauthoryear{Jaimungal and Surkov}{Jaimungal and
  Surkov}{2013}]{jaimungal2013valuing}
Jaimungal, S. and V.~Surkov (2013).
\newblock Valuing early-exercise interest-rate options with multi-factor affine
  models.
\newblock {\em International Journal of Theoretical and Applied Finance\/}~{\em
  16\/}(06), 1350034.

\bibitem[\protect\citeauthoryear{Kiesel, Schindlmayr, and Boerger}{Kiesel
  et~al.}{2009}]{kiesel2009two}
Kiesel, R., G.~Schindlmayr, and R.~H. Boerger (2009).
\newblock A two-factor model for the electricity forward market.
\newblock {\em Quantitative Finance\/}~{\em 9\/}(3), 279--287.

\bibitem[\protect\citeauthoryear{Kobari, Jaimungal, and Lawryshyn}{Kobari
  et~al.}{2014}]{kobari2014real}
Kobari, L., S.~Jaimungal, and Y.~Lawryshyn (2014).
\newblock A real options model to evaluate the effect of environmental policies
  on the oil sands rate of expansion.
\newblock {\em Energy Economics\/}~{\em 45}, 155--165.

\bibitem[\protect\citeauthoryear{McDonald and Siegel}{McDonald and
  Siegel}{1985}]{McDonaldSiegel85}
McDonald, R. and D.~Siegel (1985).
\newblock Investment and the valuation of firms when there is an option to shut
  down.
\newblock {\em International Economic Review\/}~{\em 26(2)}, 331--349.

\bibitem[\protect\citeauthoryear{McDonald and Siegel}{McDonald and
  Siegel}{1986}]{McDonaldSiegel86}
McDonald, R. and D.~Siegel (1986).
\newblock The value of waiting to invest.
\newblock {\em Quarterly Journal of Economics\/}~{\em 101\/}(4), 707--728.

\bibitem[\protect\citeauthoryear{Metcalf and Hasset}{Metcalf and
  Hasset}{1995}]{MetcalfHasset95}
Metcalf, G. and K.~Hasset (1995).
\newblock Investment under alternative return assumptions: Comparing random
  walk and mean reversion.
\newblock {\em Journal of Economic Dynamics and Control\/}~{\em 19\/}(8),
  1471--1488.

\bibitem[\protect\citeauthoryear{Pindyck}{Pindyck}{1980}]{Pindyck80}
Pindyck, R.~S. (1980).
\newblock Uncertainty and exhaustible resource markets.
\newblock {\em Journal of Political Economy\/}~{\em 88\/}(6), 1203--1225.

\bibitem[\protect\citeauthoryear{Sadowsky}{Sadowsky}{2005}]{Sadowsky2005}
Sadowsky, J.~R. (2005).
\newblock The value of learning in the product development stage: a real
  options approach.
\newblock {\em Available at SSRN 721597\/}.

\bibitem[\protect\citeauthoryear{Sarkar}{Sarkar}{2003}]{Sarkar03}
Sarkar, S. (2003).
\newblock The effect of mean reversion on investment under uncertainty.
\newblock {\em Journal of Economic Dynamics and Control\/}~{\em 28}, 377--396.

\bibitem[\protect\citeauthoryear{Taschini and Urech}{Taschini and
  Urech}{2010}]{taschini2010real}
Taschini, L. and S.~Urech (2010).
\newblock The real option to fuel switch in the presence of expected windfall
  profits under the {EU} {ETS}.
\newblock {\em The Journal of Energy Markets\/}~{\em 3\/}(2), 27--47.

\bibitem[\protect\citeauthoryear{Trigeorgis}{Trigeorgis}{1996}]{Trigeorgis1999}
Trigeorgis, L. (1996).
\newblock {\em Real Options: Managerial Flexibility and Strategy in Resource
  Allocation}.
\newblock The MIT Press.

\bibitem[\protect\citeauthoryear{Tsekrekos}{Tsekrekos}{2010}]{tsekrekos2010effect}
Tsekrekos, A.~E. (2010).
\newblock The effect of mean reversion on entry and exit decisions under
  uncertainty.
\newblock {\em Journal of Economic Dynamics and Control\/}~{\em 34\/}(4),
  725--742.

\bibitem[\protect\citeauthoryear{Weron}{Weron}{2007}]{weron2007modeling}
Weron, R. (2007).
\newblock {\em Modeling and forecasting electricity loads and prices: a
  statistical approach}, Volume 403.
\newblock John Wiley \& Sons.

\end{thebibliography}
\bibliographystyle{Chicago}

\end{document}